\documentclass[]{spie}  

\usepackage{amsmath,amsfonts,amssymb}
\usepackage{graphicx}
\usepackage[colorlinks=true, allcolors=blue]{hyperref}

\title{Identifiability in Functional Connectivity May Unintentionally Inflate Prediction Results}
\author[a]{Anton Orlichenko}
\author[a]{Gang Qu}
\author[b]{Kuan-Jui Su}
\author[b]{Anqi Liu}
\author[b]{Hui Shen}
\author[b]{Hong-Wen Deng}
\author[a]{Yu-Ping Wang}

\affil[a]{Department of Biomedical Engineering, Tulane University, New Orleans, LA, USA}
\affil[b]{School of Medicine, Tulane University, New Orleans, LA, USA}

\authorinfo{Further author information: (Send correspondence to Anton Orlichenko)\\Anton Orlichenko: E-mail: aorlichenko@tulane.edu}

\begin{document}
\maketitle

\begin{abstract}
Functional magnetic resonance (fMRI) is an invaluable tool in studying cognitive processes in vivo. Many recent studies use functional connectivity (FC), partial correlation connectivity (PC), or fMRI-derived brain networks to predict phenotypes with results that sometimes cannot be replicated. At the same time, FC can be used to identify the same subject from different scans with great accuracy. In this paper, we show a method by which one can unknowingly inflate classification results from 61\% accuracy to 86\% accuracy by treating longitudinal or contemporaneous scans of the same subject as independent data points. Using the UK Biobank dataset, we find one can achieve the same level of variance explained with 50 training subjects by exploiting identifiability as with 10,000 training subjects without double-dipping. We replicate this effect in four different datasets: the UK Biobank (UKB), the Philadelphia Neurodevelopmental Cohort (PNC), the Bipolar and Schizophrenia Network for Intermediate Phenotypes (BSNIP), and an OpenNeuro Fibromyalgia dataset (Fibro). The unintentional improvement ranges between 7\% and 25\% in the four datasets. Additionally, we find that by using dynamic functional connectivity (dFC), one can apply this method even when one is limited to a single scan per subject. One major problem is that features such as ROIs or connectivities that are reported alongside inflated results may confuse future work. This article hopes to shed light on how even minor pipeline anomalies may lead to unexpectedly superb results.
\end{abstract}


\section{Introduction}

Functional magnetic resonance is a non-invasive imaging modality that uses the blood oxygen level dependent (BOLD) signal to infer the level of neural activity in different regions of the brain. \cite{Belliveau1991FunctionalMO} fMRI has been used to localize visual processing, \cite{Cox2003-tu} attention, \cite{Coull1998-op}\cite{Pugh1996-tm} emotional processing, \cite{Phan2002-nb}\cite{6789831}\cite{Koelsch2006-ci} and language\cite{Hernandez2001-ec} to specific locations in the cortex. It has also been used to identify hemispheric dominance for, e.g., language. \cite{Szaflarski2006-bv} Functional connectivity is the Pearson correlation between the time-varying BOLD signal of different regions of the brain. \cite{Van_den_Heuvel2010-ig} It has recently been used to predict age, \cite{10002422}\cite{8666981} sex, \cite{ICER2020105444}\cite{9146335} general fluid intelligence, \cite{Qu2021EnsembleMR}\cite{9146335} pre-clinical Alzheimer's disease, \cite{MILLAR2022119228} and schizophrenia. \cite{Wang2018-iv}\cite{Rashid2016-ct} Classification based on 4D fMRI images, not FC, is also an active area of research.\cite{10043854} Naturally, the ability to predict cognition-related endophenotypes or pre-clinical disease status is an exciting avenue for translational applications.

Although fMRI offers unmatched ability to observe neural activity in vivo in human subjects, there are two questions that must be addressed when interpreting the results of predictive studies. First, are these studies meant to establish a groundwork for a clinical system such as, e.g., an AI-based breast cancer screening tool?\cite{McKinney2020-cu} If so, then these studies must be validated in a randomized trial with thousands of subjects.\cite{Salehinejad2021ARD} By contrast, fewer than 1\% of fMRI studies in 2017 and 2018 enrolled more than 100 subjects, with most recruiting less than 30.\cite{SZUCS2020117164} 
A large number of subjects is needed partly because, in the past, fMRI has faced several replicability crises.\cite{Chen2018-dd} 
For example, Bennett et al. (2010) made the case that multiple comparison correction was indispensable in fMRI by revealing emotion-associated voxels in a dead salmon.\cite{Bennett2010-gy}
It is suspicious that fMRI-based predictions of schizophrenia status achieve above 90\% accuracy with fewer than 100 training subjects,\cite{Wang2018-iv}\cite{Rashid2016-ct} whereas genome-wide association studies find single-nucleotide polymorphisms (SNPs) explain only 23\% of schizophrenia variance,\cite{Lee2012-gx} and prediction studies based on SNPs in the UK Biobank report a maximum AUC of 0.71. \cite{Bracher-Smith2022-tw} In fact, recent studies\cite{Caputi2021-vp} and many recent posters at OHBM 2023 give a classification accuracy for schizophrenia diagnosis using FC (often times cited as the best metric) of 70-80\%.\cite{Buckova-OHBM-2023}\cite{Popov-OHBM-2023}\cite{Kanyal-OHBM-2023}

If these pipelines are not meant to be introduced clinically, are they meant to provide mechanistic insights into human cognition? This is more likely to be the case, but there is sometimes a very loose interpretation of what FC actually is. For example, the UKB description of fMRI processing\cite{UKB-Brain-Imaging-Doc} makes the point that, compared to PC, FC ``has various practical and interpretational disadvantages including an inability to differentiate between directly connected nodes and nodes that are only connected via an intermediate node."\cite{Smith2012-fz} Many recent studies also implicitly assume that connectivity in the context of fMRI implies physical connections.\cite{Bastos2015-nm} 

In reality, there is no signal traveling from node to node: fMRI essentially measures blood flow,\cite{Belliveau1991FunctionalMO} and any correlation-based metric is only looking at how much the bandpass-filtered BOLD signals between two regions are in sync.\cite{Van_den_Heuvel2010-ig} Thus, at first order, fMRI is not measuring the electrical activity of neurons or the release of neurotransmitters, although reframing the problem as connectivity or graph edges may be a useful construct.\cite{9933896} On the other hand, fMRI has identified several robust characteristics of BOLD signal. In particular, it has been shown that, on average, FC intensity decreases progressing from children to young adults,\cite{Dosenbach2010-ef} and that females have greater relative intra-default mode network (DMN) connectivity compared to males.\cite{Mak2017-yj}\cite{Ficek-Tani2022-yw} FC has also shown some ability to predict race, even between different datasets,\cite{Orlichenko2023-xl} and it has been used in mechanistic studies of aggression related to olfactory stimulus. \cite{Mishor2021-ma} 

One thing that fMRI-based FC is very good at is identifying the same subject from different scans, referred to as fingerprinting.\cite{Finn2015-ft} FC can easily achieve 60\% fingerprinting accuracy,\cite{Finn2015-ft} and with post-processing, fingerprinting accuracy can become greater than 95\%.\cite{Cai2019-vp}\cite{Orlichenko2023-ad} Some studies have explicitly aimed to improve prediction of Alzheimer's disease by maximizing identifiability after processing with PCA.\cite{Svaldi2021-pz} At the same time, recent work shows that confounder elimination\cite{Hamdan2022-se} or intentional but undetectable data manipulation\cite{Rosenblatt2023-pa} can greatly improve prediction performance. 

In this work, we present a procedure by which FC-based prediction results may be unintentionally inflated. By including different scans of the same subject in both train and test sets, the machine learning algorithm learns to memorize subjects from different scans rather than to select task-specific features. Connectivities or regions that are reported in such studies may confuse other researchers when surveying the literature. Abu-Mostafa et al. (2012) gives the example of a machine learning algorithm that was able to predict exchange rate direction 52.1\% of the time, resulting in a theoretical profit of 22\% over 2 years.\cite{Learning-from-Data} In live trading, the program actually lost money. The reported problem was that the training set was normalized using the statistics of the entire cohort, and this was enough to poison the results.\cite{Learning-from-Data} If these problems show up in finance, where money is on the line, then we posit it may be prudent to look for them in scientific procedures as well.

\section{Methods}

We first present the procedure for exploiting identifiability by treating independent scans as independent subjects. Second, we describe what we mean by identifiability or fingerprinting. Third, we give a brief review of how we derive FC or dFC from 4D fMRI volumes. Finally, we list revelant characteristics of the four datasets used in this study.

\subsection{Procedure for Exploiting Identifiability}

The procedure for exploiting identifiability is simple, and example code demonstrating exploitation of identifiability is provided in the link in the footnote.\footnote{\url{https://github.com/aorliche/fc-identifiability-exploit}} When multiple longitudinal or contemporaneous scans of the same subject are available, treat these scans as independent subjects when creating training and test sets. If only one subject scan is available, use dynamic functional connectivity to create FC from multiple non-overlapping windows, and treat these FC matrices as independent subjects. In our experiments, to highlight the maximum possible gap in prediction, each subject has one scan in the training set and one scan in the test set. A random distribution of scans will achieve an accuracy somewhere between the double-dipping and legitimate results.

\subsection{Identifiability}

We define successful identification (identifiability) of a subject as a same-subject, different-scan FC pair having a higher cosine similarity (Equation~\ref{eq:cosinesim}) compared to all other scan pairs in the cohort, where $\mathbf{a}$ and $\mathbf{b}$ are vectorized subject FCs. An alternative is to use Euclidean distance as the similarity metric; the numbers we present, however, are based on cosine similarity.

\begin{equation}
    \label{eq:cosinesim}
    \text{sim}(\mathbf{a},\mathbf{b}) = \frac{\mathbf{a}^{\text{T}}\mathbf{b}}{{\lVert\mathbf{a}\rVert}_2{\lVert\mathbf{b}\rVert}_2}
\end{equation}

\begin{figure}
    \centering
    \hspace*{-1cm}\includegraphics[width=19cm]{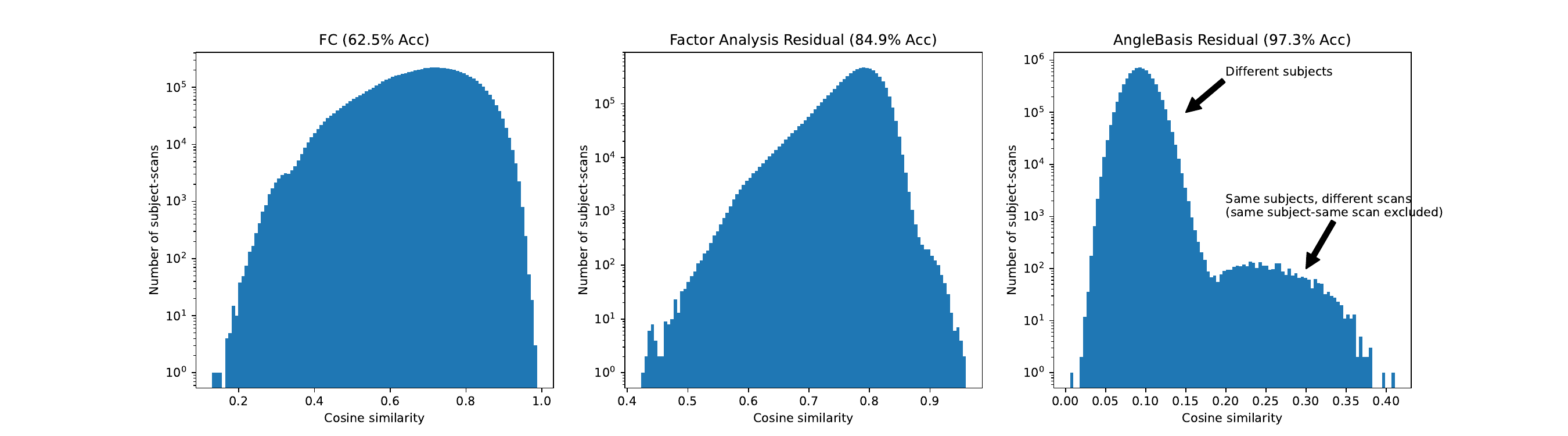}
    \caption{Demonstration of identifiability/fingerprinting with plain FC (62.5\% left) vs with FC factor analysis residual (84.9\% middle) vs with FC angle basis residual (97.3\% right). Among 1,529 subjects having 3,843 scans, same-subject, different-scan FC has the highest cosine similarity among all scan pairs 62.5\% of the time. Scans from the PNC dataset. Reproduced from Orlichenko et al. (2023).\cite{Orlichenko2023-ad}}
    \label{fig:ident}
\end{figure}

We see in Figure~\ref{fig:ident} that plain FC has 62.5\% identiability among 3,843 subjects in the PNC dataset. With preprocessing, this number can be increased to 97.3\%.\cite{Finn2015-ft}\cite{Cai2019-vp}\cite{Orlichenko2023-ad}

\subsection{Functional and Dynamic Functional Connectivity}

\begin{figure}
    \centering
    \includegraphics[width=16cm]{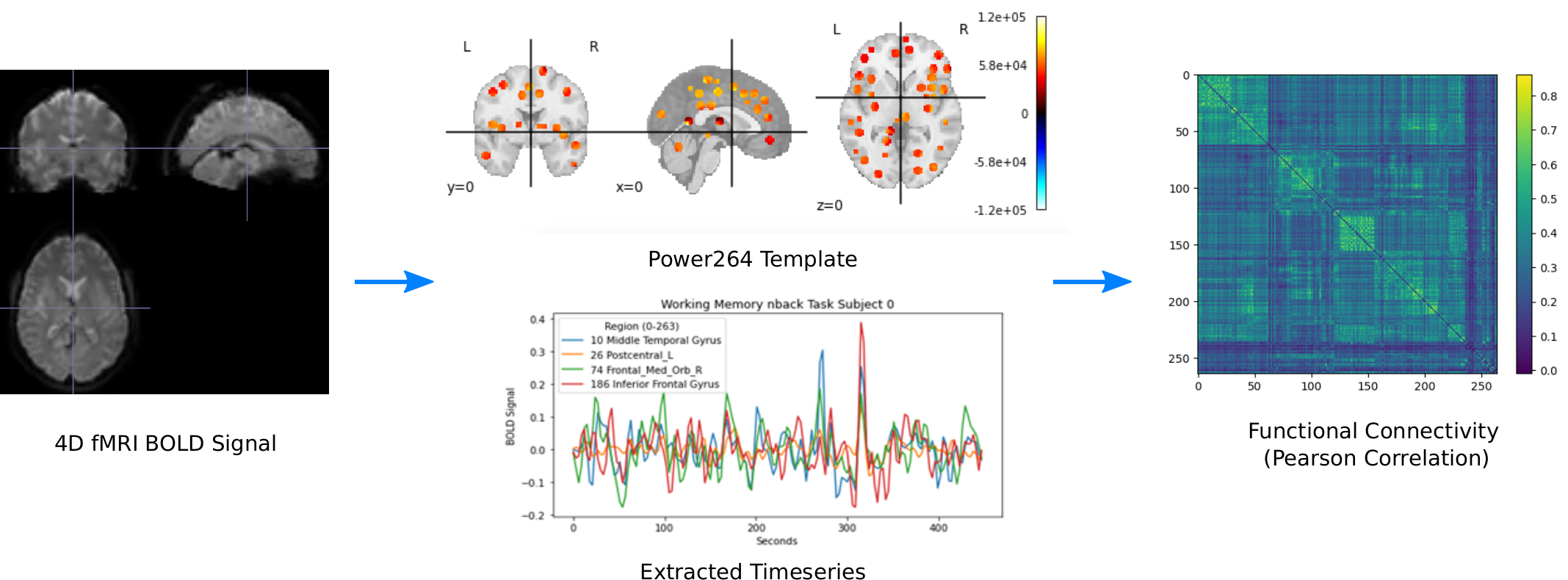}
    \caption{Illustration of the pipeline for creating FC matrices from fMRI data.}
    \label{fig:pipeline}
\end{figure}

First, we register 4D fMRI volumes into MNI space using SPM12.\footnote{\url{http://www.fil.ion.ucl.ac.uk/spm/software/spm12/}} Second, we identify regions of interest and extract the BOLD signal from those regions. These regions may be defined either using ICA\cite{Calhoun2009-fx} or a template. We use the Power264 template in this work.\cite{Power2011-fd} Third, we bandpass filter these timeseries within a 0.01 to 0.15 Hz envelope. This removes both low-frequency scanner drift and high-frequency noise as well as heartbeat and breathing signal. Finally, we calculate the Pearson correlation between the timeseries of each region to find the region-to-region FC. This symmetric matrix is reduced to the upper right triangle and vectorized. The entire procedure is illustrated in Figure~\ref{fig:pipeline}.

When no longitudinal or contemporaneous scans are available for a subject, we create multiple FC matrices from the same scan using windowing in time. Non-overlapping windows of the bandpass-filtered timeseries are used to create multiple FC matrices. In this study we use a window size of $N=50$ repetition times (TRs).

\subsection{Predictive Models}

We use simple logistic and ridge regression models for all predictive tasks. The scikit-learn implementation\cite{scikit-learn} is used in all cases.\footnote{\url{https://scikit-learn.org/stable/}} All prediction tasks are performed with an 80/20 training/test split, over 20 bootstrap iterations. The optimal hyperparameter (there is only one for both logistic and ridge regression) is chosen via grid search with grid locations at powers-of-ten intervals, i.e., on a logarithmic grid.

\subsection{Datasets}

We verify the potential of identiability to skew results in a favorable manner on four different datasets: the UK Biobank, the Philadelphia Neurodevelopmental Cohort, the Bipolar and Schizophrenia Network for Intermediate Phenotypes, and an OpenNeuro Fibromyalgia dataset.

\subsubsection{UK Biobank (UKB)}

We have processed the scans of more than 40,000 UK Biobank\cite{Sudlow2015-zq} subjects using SPM12. Of these, 2,722 subjects have two longitudinal scans, taken approximately two years apart. An additional 154 subjects have the second scan but not the first, resulting from quality control or a failure in our pipeline during pre-processing. We use the longitudinal subjects to predict age and genetic sex. We also predict age and sex on the non-longitudinal cohort in order to provide a baseline for model performance without double-dipping.

\subsubsection{Philadelphia Neurodevelopmental Cohort (PNC)}

The Philadelphia Neurodevelopmental Cohort is a dataset of 9,267 children and young adults aged 8-23 years old containing demographics, cognitive battery, questionaire responses, and SNP data.\cite{Glessner2010-xg} Among the cohort, 1,529 subjects have fMRI scans with up to 3 scanner tasks: resting state, working memory (nback), and emotion identification (emoid).\cite{Satterthwaite2014NeuroimagingOT} The data includes Wide Range Achievement Test (WRAT) scores\cite{Sayegh2014QualityOE} that have had the effects of age regressed out. It has previously been shown that the ability to predict WRAT score from FC was mostly due to the different distribution of WRAT scores among races and the ability to predict race from FC.\cite{Orlichenko2023-xl}

\subsubsection{OpenNeuro Fibromyalgia Dataset (Fibro)}

We include a 66-subject dataset of 33 female fibromyalgia patients and 33 female healthy controls from the OpenNeuro repository,\cite{Markiewicz2021-lt} study identifier ds004144.\cite{ds004144:1.0.1} Out of the entire cohort, 65 subjects have two different scans: resting state and epr. A variety of medication, demographic, and questionnaire data are available.

\subsubsection{Bipolar and Schizophrenia Network for Intermediate Phenotypes (BSNIP)}

The Bipolar and Schizophrenia Network for Intermediate Phenotypes is a large study of schizophrenia, bipolar, and schizoaffective disorder patients; relatives of patients; and healthy controls from several sites.\cite{Tamminga2014-sq} Our data contains 199 schizophrenia patients and 243 healthy controls. Patient sex is slightly skewed toward males in the schizophrenia group. Only one scan is available per subject, necessitating use of dFC in order to exploit identifiability.

\section{Results}

We present a summary of our results in Table~\ref{tab:results}, before highlighting results in each individual dataset.

\begin{table}
    \label{tab:results}
    \centering
    \begin{tabular}{|c|c|c|c|c|c|}
        \hline
         Dataset & Task & Null Model & Best Prediction & Double-Dipping & Unintentional \\
         & & & & Prediction & Improvement \\
         \hline
         UKB & Sex (Accuracy) &
         $0.528$ &
         $0.82\pm0.02$ & $0.89\pm0.006$ & $\mathbf{7\%}$ \\
         \hline
         UKB & Age (RMSE) &
         $7.68$ & $5.92\pm0.03$ & $4.97\pm0.50$ & $\mathbf{12.4\%}$ \\
         \hline
         PNC & WRAT (RMSE) & $14.6$ & $14.45\pm0.92$ & $11.0\pm0.28$ & $\mathbf{23.6\%}$ \\ 
         \hline
         Fibro & Diagnosis (Accuracy) & $0.51$ & $0.61\pm0.17$ & $0.86\pm0.038$ & $\mathbf{25\%}$ \\
         \hline
         BSNIP & Diagnosis (Accuracy) & 0.5 & $0.78\pm0.05$ & $0.89\pm0.04$ & $\mathbf{11\%}$ \\
         \hline
    \end{tabular}
    \caption{Prediction results with and without erroneous exploitation of identifiability. Best Prediction for UKB is reported for 1350 subjects in the training set, whereas for all other datasets it is reported with the maximum number of training subjects available.}
\end{table}

\subsection{UKB}

\begin{figure}
    \centering
    \hspace*{-1cm}
    \includegraphics[width=18cm]{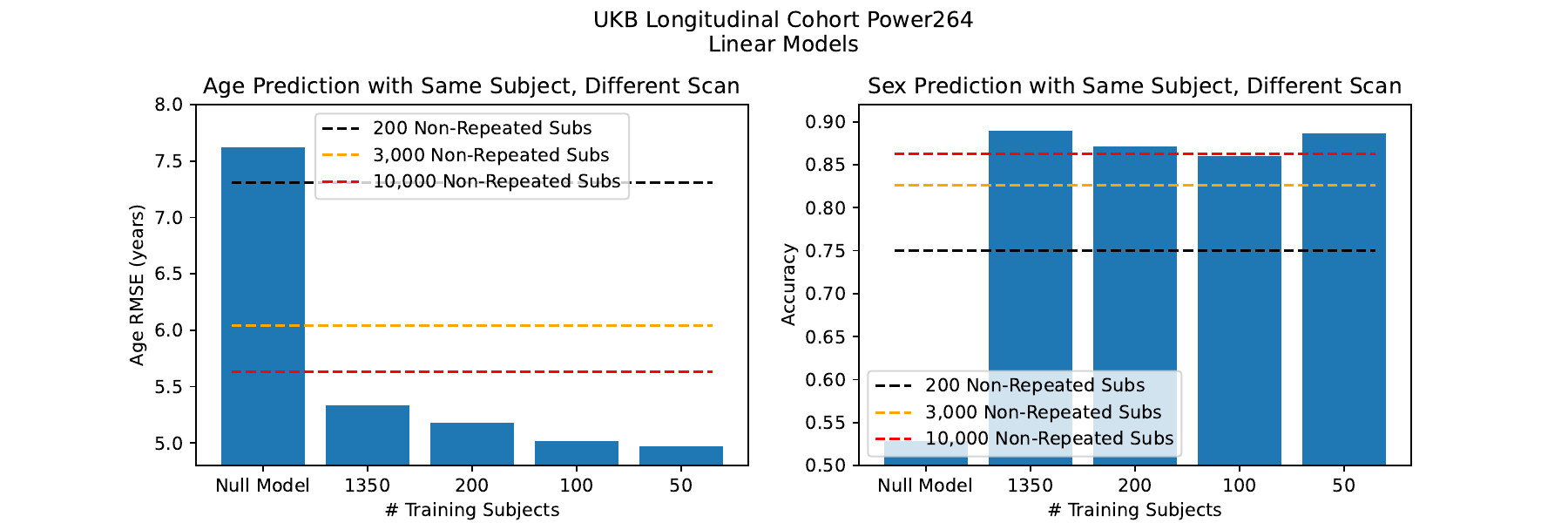}
    \caption{Prediction performance in the UKB with and without misuse of identifiability. Age prediction (left) and sex prediction (right). We find misused identifiability can lead to superb results with very small number of training subjects.}
    \label{fig:ukb}
\end{figure}

The accuracy of UKB predictions with and without unintentional identifiability enhancement (double-dipping) is shown in Figure~\ref{fig:ukb}. We find that misusing identifiability in the longitudinal cohort leads to prediction performance with 50 subjects not matched by 10,000 training subjects in the full cohort.

\subsection{PNC}

\begin{figure}
    \centering
    \includegraphics[width=14cm]{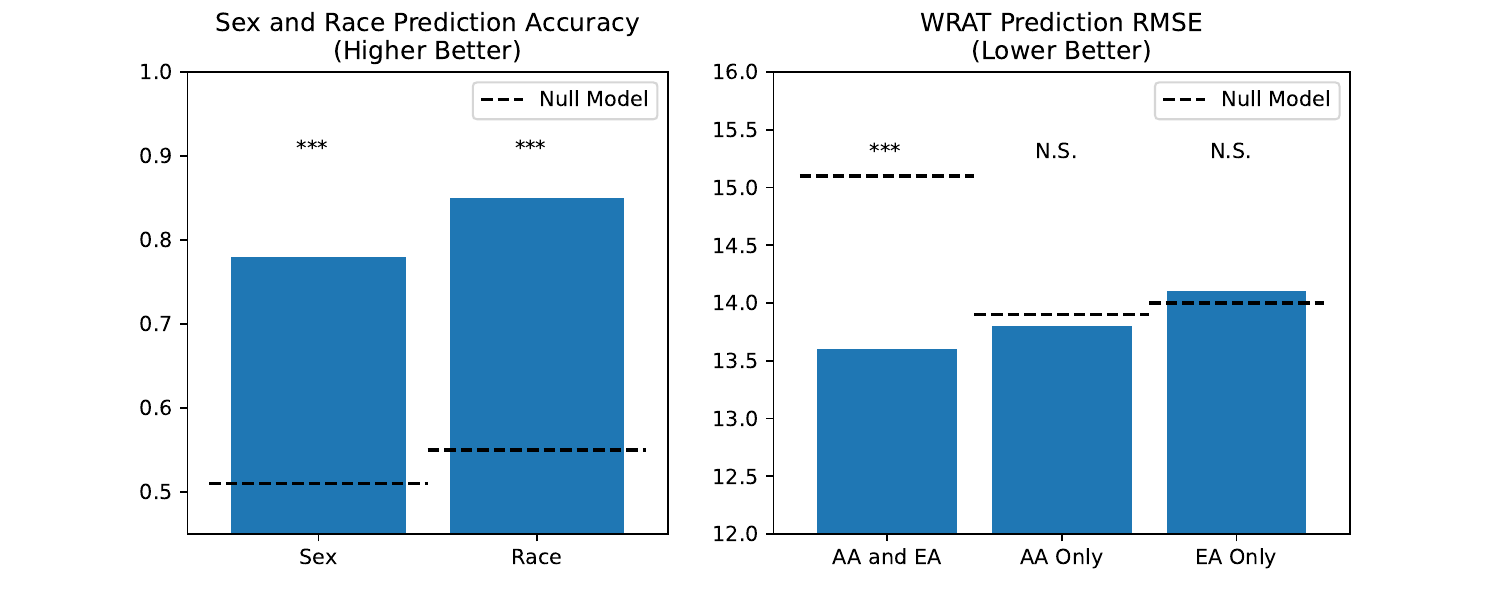}
    \includegraphics[width=14cm]{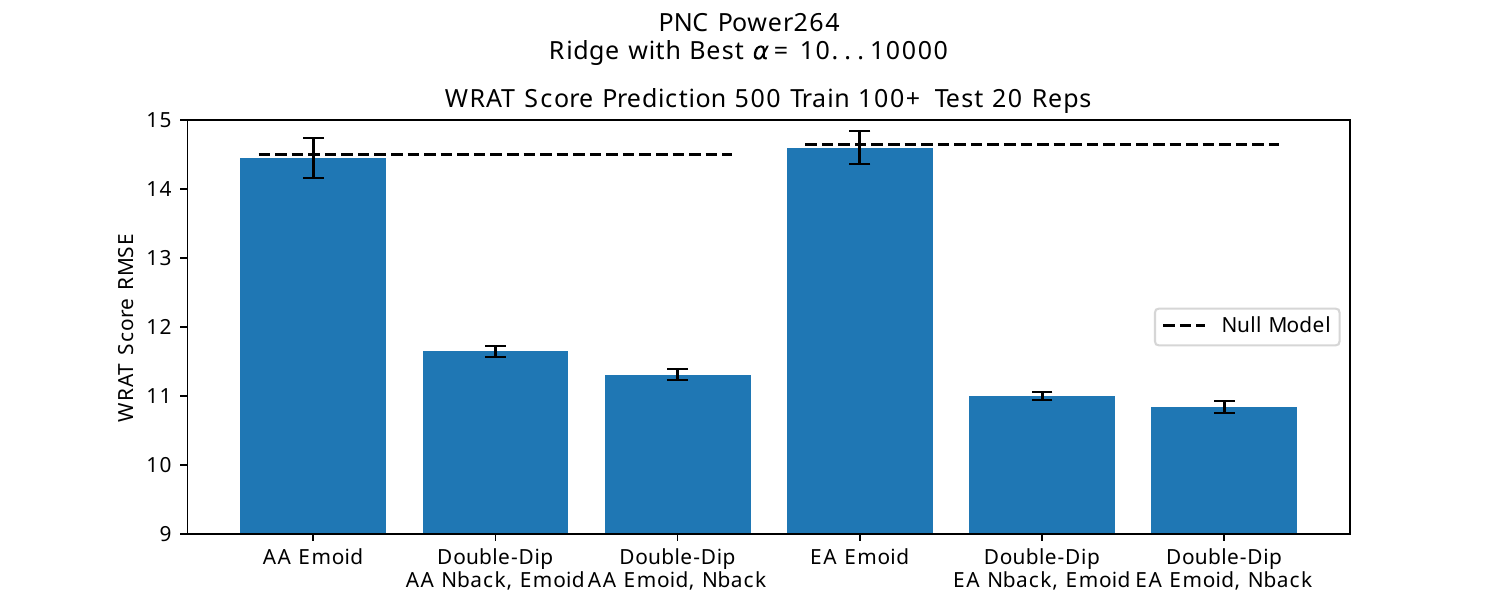}
    \caption{Incorrect attribution of FC ability to predict race as FC ability to predict achievement score (top), and the greater predictive accuracy enhancement possible by treating independent scans as independent subjects (bottom). EA refers to European Ancestry and AA refers to African Ancestry. Top graph from Orlichenko et al. (2023).\cite{Orlichenko2023-xl}}
    \label{fig:pnc}
\end{figure}

Figure~\ref{fig:pnc} (top) shows the possibility of incorrectly attributing achievement score prediction to fMRI because of a race confound.\cite{Orlichenko2023-xl} In fact, an even greater prediction accuracy can be achieved by treating independent scans as different subjects, as seen in Figure~\ref{fig:pnc} (bottom).

\subsection{Fibromyalgia}

\begin{figure}
    \centering
    \includegraphics[width=10cm]{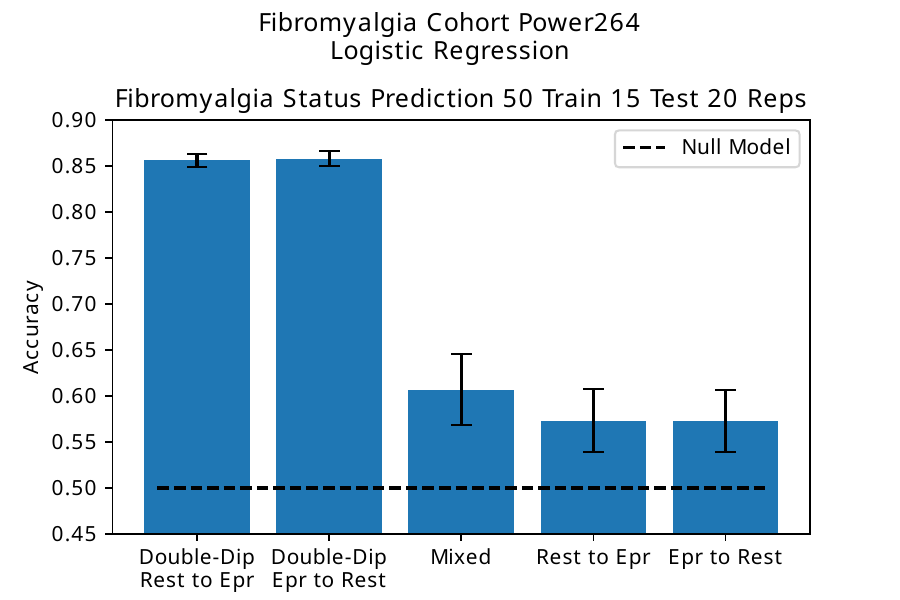}
    \caption{Prediction accuracy in the Fibromyalgia dataset using resting state scans as the training set and epr scans as the test set (and vice versa) compared to performing prediction on only one set of scans.}
    \label{fig:fibro}
\end{figure}

We see in Figure~\ref{fig:fibro} that in cohorts with a limited number of subjects, such as the Fibromyalgia dataset, the difference between proper and improper placement of scans in training and test sets in term of prediction accuracy is maximized.

\subsection{BSNIP}

\begin{figure}
    \centering
    \includegraphics[width=13cm]{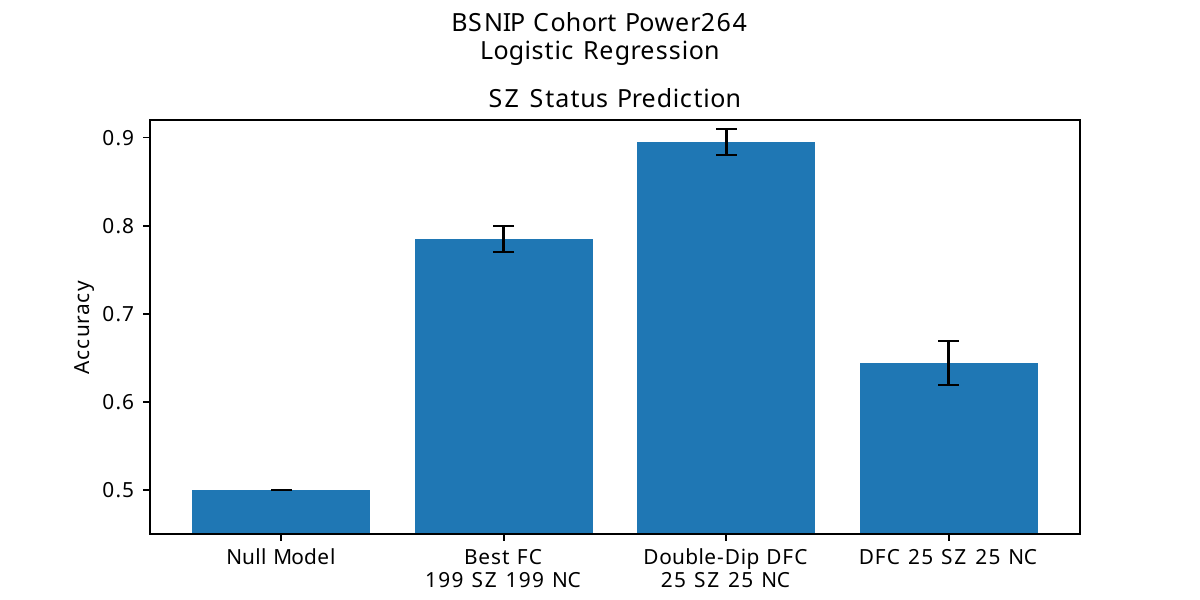}
    \caption{Schizophrenia diagnosis prediction using FC, correctly applied dFC, and dFC with using different connectivity matrices as independent subjects. We find that treating different FC matrices of the same subject as independent subjects leads to skewing of prediction results.}
    \label{fig:bsnip}
\end{figure}

As the BSNIP dataset only provides one scan per subject, identifiability enhancement must rely on the use of dynamic functional connectivity, with different windows treated as independent subjects. Identiability enhancement (Figure~\ref{fig:bsnip}) leads to predictive accuracy results in keeping with some of the larger values found in the literature.

\section{Discussion}

Various studies have examined the reproducibility of regions identified through fMRI studies.\cite{Chen2018-dd}\cite{Elliott2020-ot} To our knowledge, few studies have examined the reproducibility of, e.g., fMRI schizophrenia classification results. One multi-site study did report classification accuracies of 79.8 to 97.1\%,\cite{Salman2023-fx} the lower end of which is consistent with our own findings. Another methodology-based study found various methods to provide between 56.7 and 92.5\% classification accuracy.\cite{9740146} We use schizophrenia as an example, but any type of phenotype prediction may be used instead. While not a predictive study, van den Heuvel et al. (2017) showed that with a small number of subjects, minor manipulation of proportional thresholding can cause group differences to appear or disappear in an fMRI schizophrenia dataset.\cite{Van-den-Heuvel2017-yv} There is another effect where an artificially inflated 95\% classification accuracy existing in the literature may inhibit other researchers from publishing results that only achieve a 70-80\% accuracy in their own data.

There have been at least three recent high-profile cases of data manipulation in academia: a Harvard scientist faking data,\cite{Francesca-Gino} a Stanford scientist implicitly allowing graduate students to fake data,\cite{Marc-Tessier-Lavigne} and evidence of fake graphs in a room temperature semiconductor publication.\cite{Marel2023-fe} We are not sure how prevalent this practice is in the fMRI literature, but we think based on evidence presented earlier that some misuse may occur, especially since we provide a procedure by which one may misuse longitudinal or contemporaneous scans unintentionally. 

In regards to the UK Biobank, as more data is released, people may have a greater opportunity to reuse longitudinal data from the same individuals for prediction. Another potential unexplored effect is whether identifiability extends to family members. In this case, FC similarity due to relatedness may be mistaken for true FC-phenotype correlations.




\subsection{Potential Solutions}

The most obvious solution is not treating different scans as independent subjects and not using different dynamic FC windows as independent subjects. Otherwise, we recommend training a model on one dataset and testing on another, thus eliminating the possibility of subject memorization. Additionally, prediction results should be corroborated through different machine learning models and methods. This means that a result should only be considered valid when it is identified by several different models, not just a single newly proposed model, except with good justification. The use of very reduced feature sets (only up to 10 features per subject) may also hinder the ability of complex models to memorize identifiable subject features, even though predictive performance will almost certainly decrease. Finally, the use of a mixup model, as found in the computer science literature, may be explored as a mitigating strategy.\cite{Zhang2017-op}

\section{Conclusion}

We find that unintentional treatment of independent scans as independent subjects can greatly increase predictive accuracy. Prediction accuracy is increased by 7 to 25\% compared to the best legitimate training procedure, using a small fraction of training subjects. This highlights the importance of reproducibility studies, as well as meaningful physiological interpretations of prediction results in contrast to optimization of prediction accuracy. It would be especially helpful if machine learning studies using neuroimaging data made proposals that could be tested in an independent manner.

\section{Acknowledgements}

The authors would like acknowledge the NIH (grants R01 GM109068, R01 MH104680, R01 MH107354, P20 GM103472, R01 EB020407, R01 EB006841, R56 MH124925) and NSF (grant \#1539067) for partial funding support.

fMRI and phenotype data for the PNC dataset came from the Neurodevelopmental Genomics: Trajectories of Complex Phenotypes database of genotypes and phenotypes repository, dbGaP Study Accession ID phs000607.v3.p2. The authors would also like to thank the UK Biobank (UKB application ID 61915), the BSNIP study organizers, and OpenNeuro as well as the Fibromyalgia dataset curators for making data publicly available or available to authorized researchers.

\bibliography{identifiability} 
\bibliographystyle{spiebib} 

\end{document}